\documentclass[fleqn,usenatbib]{mnras}

\usepackage{newtxtext,newtxmath}

\usepackage[T1]{fontenc}
\usepackage{ae,aecompl}

\usepackage{graphicx}	
\usepackage{amsmath}	
\usepackage{amssymb}	
\usepackage{changes}
\usepackage{relsize}

\def\apj{ApJ}
\def\apjl{ApJL}
\def\apjs{ApJ Suppl. Ser.}

\def\aap{A\&A}

\def\araa{ARAA}
\def\mnras{MNRAS}

\def\pasj{PASJ}

\def\sovast{Sov. Astron.}

\def\beq#1{\begin{equation}\label{#1}}
\def\eeq{\end{equation}}
\def\beqa#1{\begin{eqnarray}\label{#1}}
\def\eeqa{\end{eqnarray}}
\def\eq#1{eq.~(\ref{#1})}
\def\Eq#1{Eq.~(\ref{#1})} 

\def\myfrac#1#2{\left(\frac{#1}{#2}\right)}

\def\comment#1{\relax}

\newcommand{\msy}{M_\odot\, \mbox{yr}^{-1}}

\newcommand{\kms}{\:\mbox{km\,s}$^{-1}$ }

\title[Mass ratio in SS433]{Mass ratio in SS433 revisited}

\author[A.M. Cherepashchuk et al.]{
A.M. Cherepashchuk$^{1}$\thanks{E-mail: cherepashchuk@gmail.com},
K.A. Postnov$^{1,2}$\thanks{E-mail: pk@sai.msu.ru}
and 
A.A. Belinski$^1$\thanks{E-mail: aleks@sai.msu.ru}
\\
$^{1}$ Sternberg Astronomical Institute, M.V. Lomonosov Moscow State University, 13, Universitetskij pr., 119234, Moscow, Russia\\
$^{2}$ Kazan Federal University, Kremlevskaya 18, 420008 Kazan, Russia
}

\date{Accepted XXX. Received YYY; in original form ZZZ}

\pubyear{2019}

\begin{document}
\label{firstpage}
\pagerange{\pageref{firstpage}--\pageref{lastpage}}
\maketitle

\begin{abstract}
We revisit the determination of binary mass ratio in the Galactic microquasar SS433 based on recent GRAVITY VLTI measurements of mass and angular momentum outflow through a circumbinary disc. The new observations combined with the constancy of the binary orbital period over $\sim 30$ yrs confirm that the mass ratio in SS433 is $q=M_\mathrm{x}/M_\mathrm{v}\gtrsim 0.6$. For the assumed optical star mass $M_\mathrm{v}$ ranging from $\sim 8$ to 15 $M_\odot$ such a mass ratio suggests a low limit of the compact object mass of $M_\mathrm{x}\sim 5-9 M_\odot$, placing the compact object in SS433 as a stellar-mass black hole.  

\end{abstract}




\section{Introduction}

Long-term optical, X-ray and radio studies of the microquasar SS433 have suggested that this unique Galactic object is a high-mass eclipsing X-ray binary at advanced evolutionary stage  \citep{1979ApJ...233L..63M,1981MNRAS.194..761C,2004ASPRv..12....1F}. In SS433, a supercritical accretion disc and relativistic jets precessing with a period of $P_\mathrm{prec}\simeq 162^\mathrm{d}.3$ are observed. A recent analysis of the 40-years optical data on  Doppler shifts of moving $H_\alpha$ emission lines  \citep{2018ARep...62..747C} revealed that the precessional, nutational and orbital periods of SS433 are on average stable over 40 years. The SS433 binary system is completing the secondary mass transfer stage after which a WR+C binary should be formed (C here means the relativistic compact object). However, at odds with predictions of the standard theory of evolution of massive close binaries \citep{MT1988}, no common envelope has been formed in the system, and the excess of mass, energy and angular momentum is carried away from the binary system by a powerful stellar wind ($\dot M\approx 10^{-4} \msy$, $v\simeq 1500$ \kms) launched from the precessing accretion disc, as well as by relativistic jets ($\dot M\approx 10^{-6}-10^{-7} \msy$, $v\simeq 0.26 c$). 

Recently, important details of the high-mass close binary evolution have been recognized enabling us to better understand this unique feature of SS433 \citep{2017MNRAS.471.4256V,2015MNRAS.449.4415P,2017MNRAS.465.2092P}. According to \cite{2017MNRAS.471.4256V}, the common envelope in a massive X-ray binary can form if the component mass ratio  $q=M_\mathrm{x}/M_\mathrm{v}\lesssim 0.29$ ($M_\mathrm{x}$ and $M_\mathrm{v}$ is the mass of the relativistic and optical component, respectively). In the opposite case, $q\gtrsim 0.29$, the common envelope is not formed and the binary system evolves as a semi-detached one with a steady mass transfer from the Roche-lobe filling optical star forming a supercritical accretion disc with powerful stellar wind outflow, like in SS433. 

The stability of mass transfer through the vicinity of the inner Lagrangian point L$_1$ was investigated in several papers (see, for example, \cite{2015MNRAS.449.4415P,2017MNRAS.465.2092P}). It was shown that for massive donors with radiative envelopes in high-mass X-ray binaries and provided that $M_\mathrm{x}\gtrsim 0.29 M_\mathrm{v}$, the mass transfer through L$_1$ is stable. In this case, the optical star can remain for a long time in the state of significant overfilling of its Roche lobe, and the matter from the star can be lost through both inner L$_1$ and outer L$_2$ Lagrangian points. The effective photosphere radius of the optical star can significantly exceed that of its mean Roche lobe. In the case of SS433, this enables an anomalously large width of the X-ray (2-10 keV) eclipse of relativistic jets \citep{1989PASJ...41..491K, 1996PASJ...48..619K} to be explained with a relatively large binary mass ratio $q$.   
Thus, determination of the binary mass ratio in SS433 is very important to understand the evolution of this unique Galactic system.

A surprising feature of SS433 is the stability of the orbital binary period (see \cite{2004ASPRv..12....1F}, \cite{2018MNRAS.479.4844C} (Paper I)), which suggests a high binary mass ratio $q\gtrsim 0.6$. Recent spectrophotometric and astrometric observations of SS433 by the GRAVITY VLT interferometer  with an angular resolution of better than one milliarcsecond \citep{2018arXiv181112558W} enabled sky mapping of the region of the double-peak Brackett Br$-\gamma$ emission line in the IR-spectrum of the system. These direct observations proved, for the first time, that these emission lines are not generated in the accretion disc around the compact object but are produced in the circumbinary disc-like shell around the binary system. It is also proved that the broad component of these double-peak emission lines in SS433 is produced by the stellar wind outflow from the supercritical accretion disc. Therefore, the conclusion of our Paper I, in which a complex model of the formation of Brackett emission line profiles was suggested (double-peak component -- in the circumbinary shell, broad lines -- in the supercritical accretion disc wind) is now supported by direct observations.

According to \cite{2018arXiv181112558W}, the circumbinary shell around SS433 demonstrates a super-Keplerian motion with a significant radial velocity component, apparently suggesting a replenishment of the rotating shell by matter from the supercritical accretion disc wind outflow with a velocity of $\sim 1000$ \kms.

In Paper I, we estimated the binary mass ratio in SS433 $q\gtrsim 0.6$ from the observed constancy of the binary orbital period using the model of isotropic mass re-emission for the supecritical accretion disc wind and assuming an additional mass loss from the system through the outer Lagrangian point L$_2$. As the geometry and structure of Brackett emission line region is now established from the GRAVITY VLTI observations \citep{2018arXiv181112558W}, we are able to obtain a more reliable binary mass ratio estimate in SS433 from the observed constancy of the binary orbital period: $q\gtrsim 0.6$. This new estimate is consistent with conclusions of \cite{2017MNRAS.471.4256V} on the absence of common envelope in SS433 and confirms the status of SS433 as a semi-detached high-mass X-ray binary with stable mass transfer onto a massive compact object. Note that a high mass of the compact object in SS433 was also independently obtained in the recent paper by \cite{2018A&A...619L...4B}.  

When estimating the mass of the relativistic object in SS433 one should bear in mind that the distance to the system is well known, $d=5.5$~kpc \citep{2004ApJ...616L.159B}, and the total absorption in the direction to SS433 in the optical is $A_\mathrm{v}\simeq 8^\mathrm{m}$ \citep{1982SvA....26..697C,1984ARA&A..22..507M}. Using this information, from the visible stellar magnitude of SS433 and by fixing the fraction of the optical emission from the supercritical accretion disc ($\sim 0.6-0.7$, \cite{2008ApJ...676L..37H}), it is possible to estimate the mass of the optical component, which, according to \cite{2011PZ.....31....5G}, falls within the range $M_\mathrm{v}=8.3-12.5 M_\odot$, corresponding to the spectral class A4I-A8I. An upper mass limit of the optical component in SS433 can be set at $M_\mathrm{V}=15 M_\odot$. Therefore, the binary mass ratio $q>0.6$ implies the mass of the relativistic compact star to be $M_\mathrm{x}=qM_\mathrm{v}>(5-9)\,M_\odot$, i.e. the compact star in SS433 must be a black hole.  

\section{A new estimate of the binary mass ratio in SS433}

Like in Paper I, consider the following model of mass flows in SS433. We will assume a circular binary with orbital separation $a$. Let the mass-loss rate from the optical star $M_\mathrm{v}$ be $\dot M_\mathrm{v}$, the total mass of the system be $M=M_\mathrm{v}+M_\mathrm{x}=M_\mathrm{v}(1+q)$, the mass accreted by the compact object be 0, the mass outflow from the supercritical accretion disc via the Jeans mode (i.e. with the specific orbital angular momentum of the accretor) be $\beta\dot M_\mathrm{v}$, $\beta\le 1$, and the mass outflow from the system in a circumbinary shell be $(1-\beta)\dot M_\mathrm{v}$. Then from the angular momentum conservation we obtain the equation for the change of the binary separation (see Paper I):
\beq{e:adota}
\frac{\dot a}{a}=-2\left(1-\frac{1}{2}\frac{M_\mathrm{v}}{M}\right)\frac{\dot M_\mathrm{v}}{M_\mathrm{v}}+2\beta\frac{M_\mathrm{v}}{M_\mathrm{x}}\frac{\dot M_\mathrm{v}}{M}+2(1-\beta)\frac{\left(\frac{dJ}{dt}\right)_\mathrm{out}}{J}\,,
\eeq
where $J=(M_\mathrm{x}M_\mathrm{v}/M)\sqrt{GMa}$ is the orbital angular momentum of the binary.
Conservative mass transfer corresponds to $\beta=0$ and $dJ=0$, and in the non-conservative case $(dJ/dt)_\mathrm{out}/J$ specifies the additional orbital angular momentum loss by the mass escaping the binary in the outer shell.

In Paper I, we have assumed a model for the term $(dJ/dt)_\mathrm{out}/J$ with mass-loss carrying the specific angular momentum corresponding to that at the outer Lagrangian point L$_2$ (see, e.g., \cite{1997A&A...327..620S}). New measurements \citep{2018arXiv181112558W} enable us to specify this term directly from observations. Indeed, this term can be represented in the form 
\beq{e:jout}
\left(\frac{dJ}{dt}\right)_\mathrm{out}=(1-\beta)\dot M_\mathrm{v}v_\phi(R_\mathrm{out})R_\mathrm{out}\,,
\eeq
where $v_\phi(R_\mathrm{out})$ is the tangential component of the outflow velocity at the outer radius of the circumbinary shell $R_\mathrm{out}$. Therefore, the last term in \Eq{e:adota} takes the form:
\beq{e:joutj}
2(1-\beta)\frac{\left(\frac{dJ}{dt}\right)_\mathrm{out}}{J}=
2\frac{\dot M_\mathrm{v}}{M_\mathrm{v}}
\frac{M}{M_\mathrm{x}}
\frac{(1-\beta)v_\phi(R_\mathrm{out})R_\mathrm{out}\omega^{1/3}}{(GM)^{2/3}}\,,
\eeq
where $\omega =2\piup/P_\mathrm{b}$ is the orbital angular frequency and we have used the 3rd Kepler's law to express the orbital separation through $\omega$.

\begin{figure}
	\includegraphics[width=\columnwidth]{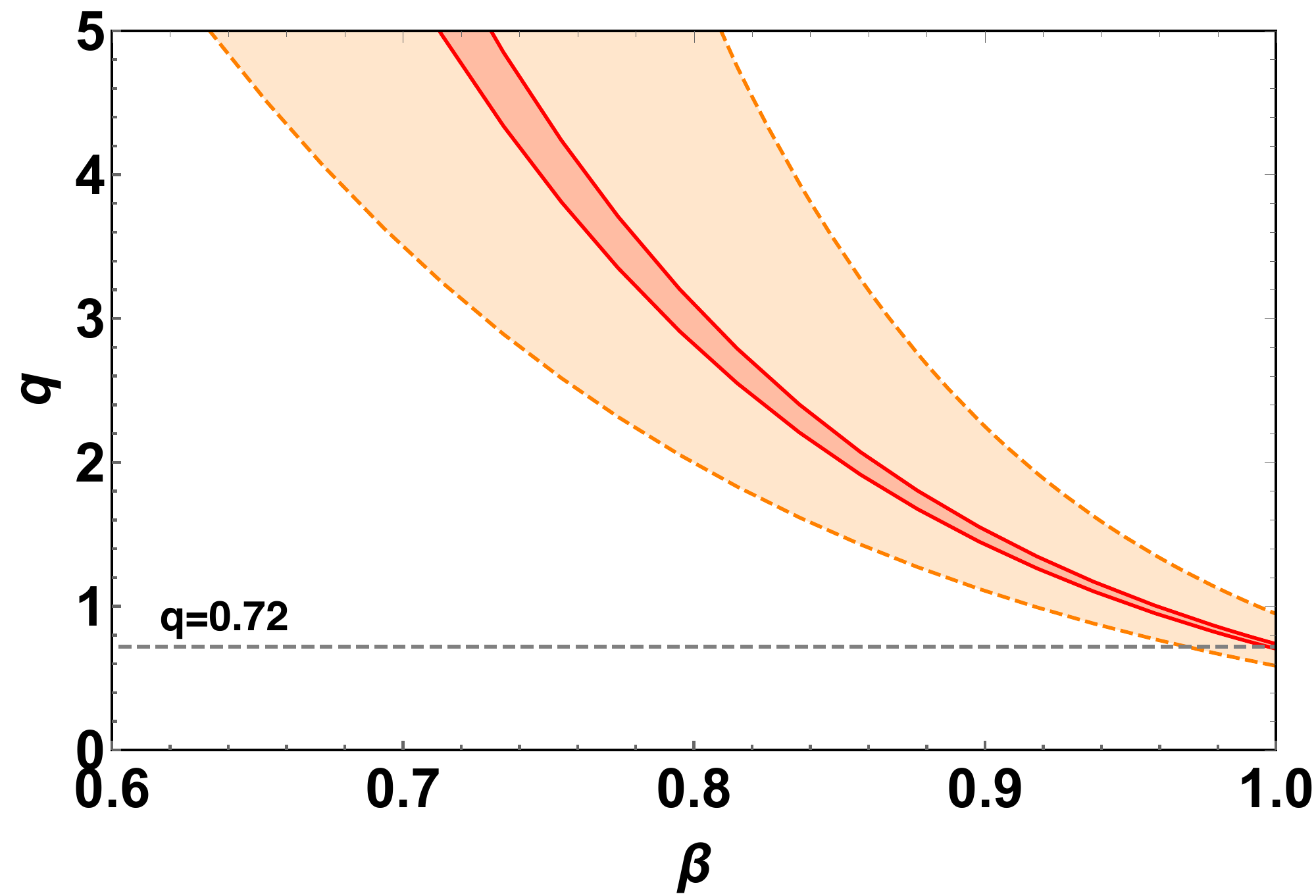}
    \caption{Constraints on the mass ratio of SS433 for different values of the mass-loss rate $\dot M_\mathrm{v}=10^{-4}$ and $10^{-5} M_\odot$ yr$^{-1}$ (intervals restricted by the solid and dashed lines, respectively), for the fiducial value of the dimensionless parameter $K=5$. The dashed horizontal line shows the solution of the quadratic equation $3q^2+2q-1=0$ corresponding to $\dot P_\mathrm{b}=0$ at $\beta=1$ (only Jeans mode mass outflow) independently of the mass-loss rate.}
    \label{f:f1}
\end{figure}

\begin{figure}
	\includegraphics[width=\columnwidth]{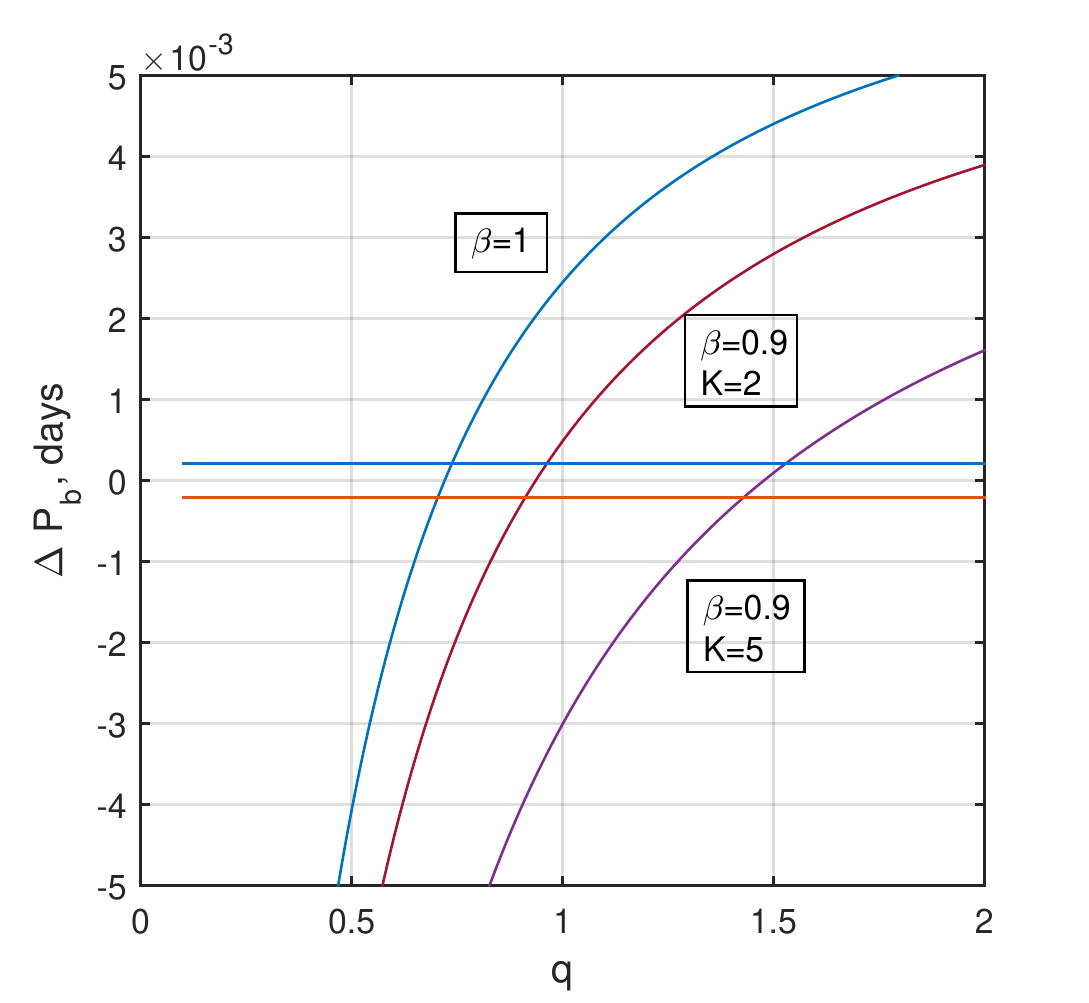}
    \caption{Expected change in the orbital period of SS433  (in days) as a function of the mass ratio $q$ (\Eq{e:dotPP}) over a 30-yr period. Shown is the limiting case of purely Jeans mode outflow ($\beta =1$) and the case of additional 10\% mass-loss ($\beta =0.9$) via the circumbinary shell parametrized by the parameter $K=5$ and $K=2$  (\Eq{e:Knum}) for the assumed total mass-loss rate $\dot M_\mathrm{V}=10^{-4} M_\odot$ yr$^{-1}$. Two parallel lines show the boundaries $\Delta P_\mathrm{b}=\pm 3\sigma_P=0^\mathrm{d}.00021$ derived from observations.}
    \label{f:f2}
\end{figure}

Changing from $\dot a/a$ to $\dot P_b/P_b$ with the help of the 3rd Kepler's law and substituting $M_\mathrm{x}=qM_\mathrm{v}$, $M=M_\mathrm{v}(1+q)$, we ultimately arrive at the equation for the fractional change of the binary orbital period (cf. Eq.(7) in Paper I):
\beq{e:dotPP}
\frac{\dot P_b}{P_b}=-\frac{\dot M_\mathrm{v}}{M_\mathrm{v}}\frac{3q^2+2q-3\beta -3K(1-\beta)(1+q)^{5/3}}{q(1+q)}\,.
\eeq
Here the dimensionless coefficient $K$ specifying the angular momentum loss via circumbinary disc reads
\beq{e:K}
K=\frac{v_\phi(R_\mathrm{out})R_\mathrm{out}}{(GM_\mathrm{v})^{2/3}}\myfrac{2\piup}{P_b}^{1/3}\,.
\eeq

Taking the orbital period $P_b=13^\mathrm{d}.1$ and normalizing $v_\phi$ and $R_\mathrm{out}$ to the characteristic values $220$ \kms and 0.7 mas (see Table 3 in \cite{2018arXiv181112558W}, the 'outflow' model favoured by the authors), we obtain 
\beq{e:Knum}
K\approx 5.1 \myfrac{v_\phi(R_\mathrm{out})}{220\,\mbox{km\,s}^{-1}}\myfrac{R_\mathrm{out}}{0.7\mbox{mas}}
\myfrac{M_\mathrm{v}}{15\,M_\odot}^{-2/3}\,.
\eeq

The large value of $K$ (cf. $x\simeq 1$ in Eq. (7) of Paper 1) is very unusual. Indeed, taking at face value, this suggests a strong increase of the specific angular momentum in the outflow compared to the orbital value. The physical mechanism for that, as the authors \cite{2018arXiv181112558W} stress, remains unknown. Moreover, the circumbinary outflow in SS433 is observed to be highly variable. Nevertheless, in our analysis below we use $K=5$ as a fiducial value.  

Acting exactly in the same way as in Paper I, from the observed stability of the orbital period of SS433 over $\Delta t\sim 28$ years, $|\dot P_b|\le 3\times \sigma_P/\Delta t$, where $\sigma_P=0^\mathrm{d}.00007$ is the uncertainty in the SS433 orbital period determination  \citep{2011PZ.....31....5G},  and assuming the optical star mass $M_\mathrm{v}=15 M_\odot$ with the mass-loss rate $\dot M_\mathrm{v}=10^{-4} M_\odot$~yr$^{-1}$, in Figure \ref{f:f1} we plot the constraints on the binary mass ratio in SS433 $q$ for $K=5$, as a function of the parameter $\beta$. 

Clearly, even a tiny fraction of the total mass loss via the circumbinary shell, $1-\beta\simeq 0.01$, places a mass ratio constraint $q\gtrsim 0.6$ in SS433. In view of high sensitivity of this estimate to the amount of matter outflow through the circumbinary shell, it is important to independently evaluate $(1-\beta)\dot M_\mathrm{v}$. For example, the estimate of the disc-like mass outflow rate from radio observations \citep{2001ApJ...562L..79B} suggests $\dot M_\mathrm{out}\sim 0.18\times 10^{-4} \msy$ (see their Eq. 1) for the source distance $d=5.5$~kpc and the outflow radial velocity at the radio optically thick distance $\sim 10^{15}$~cm of about $\sim 230\times (3.5/60)\simeq 14$~\kms (assuming the law $1/r$ for the outflow velocity as in  \cite{2018arXiv181112558W}). Therefore, a few percents of the total mass loss rate from the system are quite possibly being ejected via the circumbinary shell outflow.  

In Figure \ref{f:f2} we plot the expected change in the orbital period of SS433  (in days) over a 30-yr time interval as a function of the mass ratio $q$ calculated directly by formula \eq{e:dotPP} for the limiting case of 
purely Jeans mode outflow ($\beta =1$) and the case of additional 10\% mass-loss ($\beta =0.9$) via the circumbinary shell parametrized by the angular momentum loss parameter $K=5$ and $K=2$ (see \Eq{e:Knum}). Two parallel lines show the boundaries $\Delta P_\mathrm{b}=\pm 3\sigma_P=0^\mathrm{d}.00021$. Clearly, in all cases the binary mass ratio cannot be smaller than $q_\mathrm{min}\sim 0.6$. 

\begin{figure}
	\includegraphics[width=\columnwidth]{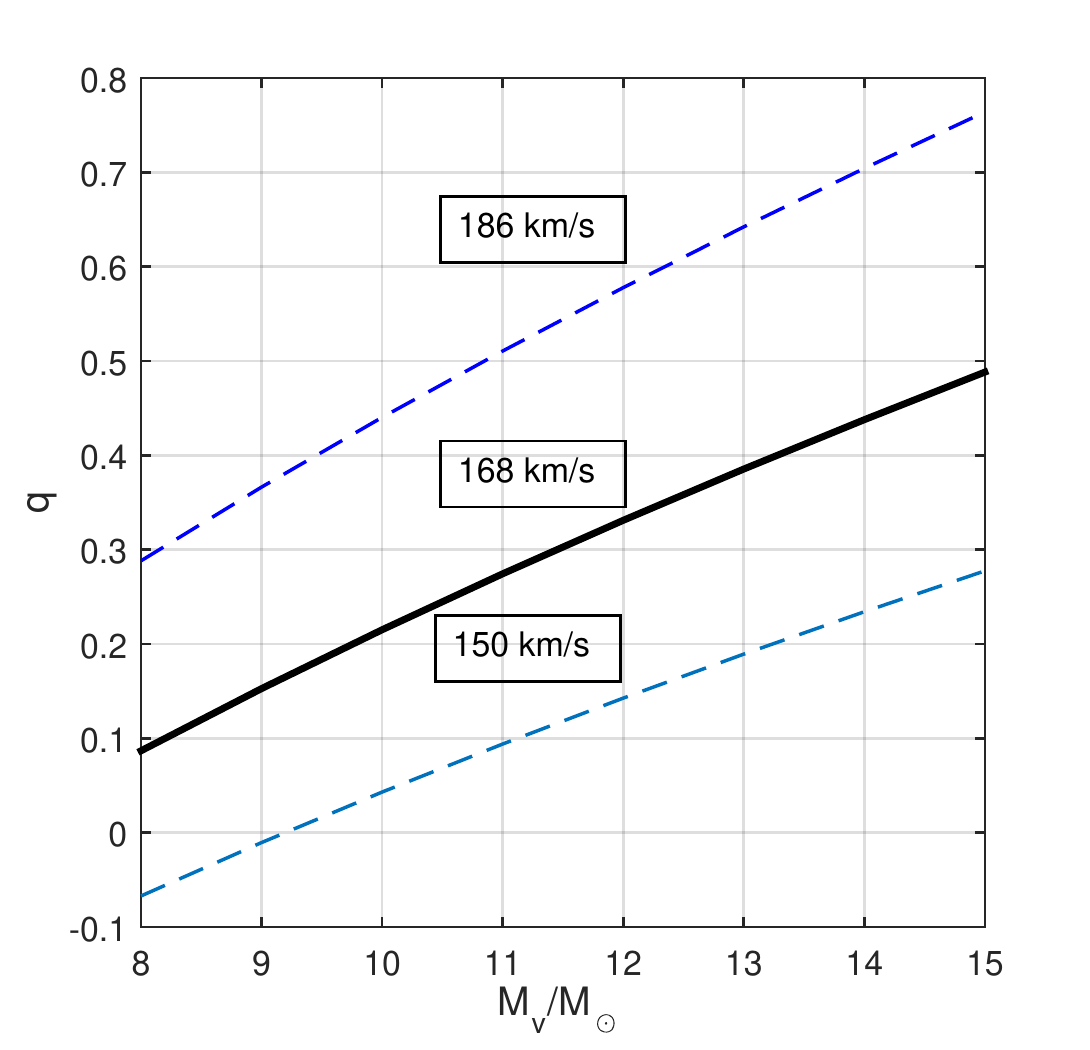}
    \caption{Binary mass ratio $q$ corresponding to the radial orbital velocity of the compact object (\Eq{e:vorb}) as required to reproduce observed HeII velocity amplitude $V_\mathrm{HeII}=168\pm 18$\kms. }
    \label{f:f3}
\end{figure}

\section{Discussion and conclusion}

In the present paper, we have re-analyzed the mass ratio constraints in SS433 imposed by the observed stability of the binary orbital period $P_\mathrm{b}$ over $\sim 30$~yr period of observations \citep{2011PZ.....31....5G} with taking into account the new GRAVITY VLTI result on the observed mass-loss through the circumbinary shell in this system \citep{2018arXiv181112558W}. This analysis strengthens our previous conclusion (see Paper I)  that the binary mass ratio $q=M_\mathrm{x}/M_\mathrm{v}$ in SS433 is likely to be $q\gtrsim 0.6$. The result is very sensitive to the (unknown) fraction of the mass-loss from the system via the circumbinary shell,
the factor $(1-\beta)$ in \Eq{e:dotPP}. Extant measurements carried out at different epochs suggest strong variability of both the mass-loss rate and geometry of the outflow \citep{2017A&A...602L..11G,2018arXiv181112558W}.  In spite of these uncertainties, even with a tiny or zero average mass-loss rate via circumbinary disc, the observed orbital period stability over about 30 years strongly points to a high binary mass ratio. It is also important to stress that irrespective of details, the angular momentum transfer from the binary system through a  circumbinary shell or a mass-loss through L$_2$ point (or both) only \textit{increase} the mass ratio estimate. 

An independent constraint on the binary mass ratio can be obtained from the observed radial velocity amplitude of HeII emission lines.
The radial orbital velocity of the compact object in SS433 for the mean binary inclination $i=79^\circ$ \citep{2018ARep...62..747C}  reads
\beq{e:vorb}
V_\mathrm{x}\sin i\approx 219 [\mathrm{km\,s}^{-1}]
\left(\frac{M_\mathrm{v}}
{15\,M_\odot}\right)^{1/3}
(1+q)^{-2/3}
\,.
\eeq
The observed radial velocity variations as measured by HeII emission lines, $V_\mathrm{HeII}= 168\pm 18$~\kms \citep{2004ApJ...615..422H}, if interpreted as being purely due to the orbital motion of the compact object, requires the binary mass ratio $q$ as shown in Fig. \ref{f:f3}. Clearly, within the measurement errors, the values $q\gtrsim 0.6$ inferred above can be accommodated for the optical component mass $M_\mathrm{V}\gtrsim 12 M_\odot$. 
For smaller masses, a controversy with the apparent lack of change of the binary orbital period would arise, even for insignificant mass loss through the circumbinary disk ($1-\beta\simeq 0$). However, one should be careful in using HeII emission lines as a proxy for the orbital motion of the compact object because of their complicated formation in the inhomogeneous wind outflow (see, e.g., discussion and references in \cite{2017ApJ...841...79R}). 

The mass of the optical component in SS433 remains also debatable. Indeed, the interstellar absorption in the direction to SS433, $A_\mathrm{V}\approx 8^m$, is very large and is estimated with significant errors, and the contribution of the accretion disc to the total optical luminosity of SS433 is also derived spectrophotometrically with significant uncertainty. Thus the mass estimate of the optical component from its bolometric luminosity is not very reliable. In addition, as noted by \cite{2010A&A...521A..81B,2011A&A...531A.107B}, the absorption lines in SS433 spectrum can arise in the circumstellar shell and not in the optical star atmosphere, therefore the spectral classification and mass of the optical star can be questionable.

We also note that the derived lower limit on the mass ratio in SS433 $q\gtrsim 0.6$  pointing to a higher mass of the optical star (see Fig. \ref{f:f3}) contradicts to the optical star's radial velocity curve based on its classification as an A supergiant \citep{2004ApJ...615..422H,2010ApJ...709.1374K} and suggesting $q\lesssim 0.3$. Clearly, here more studies are required.

We stress that the observed highly unstable character of the outflow parameters in SS433 (see e.g. \cite{2011MNRAS.417.2401B}) does not allow us to infer a more accurate estimate. Further observations can be helpful to put more precise limits on the binary mass ratio in this system. Nevertheless, even at the present stage of knowledge of the system parameters, the obtained binary mass ratio lower limit $q\gtrsim 0.6$ seems to be quite robust, placing the compact object in SS433 in the realm of Galactic stellar-mass black holes. 

\section*{Acknowledgements}

We thank the anonymous referee for useful notes and suggestions.
The work of AMCh is supported by the RSF grant 17-12-01241 (analysis of constraints from X-ray eclipses and circumbinary shell).
The work of KAP and AAB (derivation of the constraints on $q$ from observations, Section 2) is supported by the Program of development of M.V. Lomonosov Moscow State University (Leading Scientific School 'Physics of stars, relativistic objects and galaxies').




\bibliographystyle{mnras}
\bibliography{ss433} 

\begin{thebibliography}{}
\makeatletter
\relax
\def\mn@urlcharsother{\let\do\@makeother \do\$\do\&\do\#\do\^\do\_\do\%\do\~}
\def\mn@doi{\begingroup\mn@urlcharsother \@ifnextchar [ {\mn@doi@}
  {\mn@doi@[]}}
\def\mn@doi@[#1]#2{\def\@tempa{#1}\ifx\@tempa\@empty \href
  {http://dx.doi.org/#2} {doi:#2}\else \href {http://dx.doi.org/#2} {#1}\fi
  \endgroup}
\def\mn@eprint#1#2{\mn@eprint@#1:#2::\@nil}
\def\mn@eprint@arXiv#1{\href {http://arxiv.org/abs/#1} {{\tt arXiv:#1}}}
\def\mn@eprint@dblp#1{\href {http://dblp.uni-trier.de/rec/bibtex/#1.xml}
  {dblp:#1}}
\def\mn@eprint@#1:#2:#3:#4\@nil{\def\@tempa {#1}\def\@tempb {#2}\def\@tempc
  {#3}\ifx \@tempc \@empty \let \@tempc \@tempb \let \@tempb \@tempa \fi \ifx
  \@tempb \@empty \def\@tempb {arXiv}\fi \@ifundefined
  {mn@eprint@\@tempb}{\@tempb:\@tempc}{\expandafter \expandafter \csname
  mn@eprint@\@tempb\endcsname \expandafter{\@tempc}}}

\bibitem[\protect\citeauthoryear{{Blundell} \& {Bowler}}{{Blundell} \&
  {Bowler}}{2004}]{2004ApJ...616L.159B}
{Blundell} K.~M.,  {Bowler} M.~G.,  2004, \mn@doi [\apjl] {10.1086/426542},
  \href {http://adsabs.harvard.edu/abs/2004ApJ...616L.159B} {616, L159}

\bibitem[\protect\citeauthoryear{{Blundell}, {Mioduszewski}, {Muxlow},
  {Podsiadlowski}  \& {Rupen}}{{Blundell} et~al.}{2001}]{2001ApJ...562L..79B}
{Blundell} K.~M.,  {Mioduszewski} A.~J.,  {Muxlow} T.~W.~B.,  {Podsiadlowski}
  P.,   {Rupen} M.~P.,  2001, \mn@doi [\apjl] {10.1086/324573}, \href
  {http://adsabs.harvard.edu/abs/2001ApJ...562L..79B} {562, L79}

\bibitem[\protect\citeauthoryear{{Blundell}, {Schmidtobreick}  \&
  {Trushkin}}{{Blundell} et~al.}{2011}]{2011MNRAS.417.2401B}
{Blundell} K.~M.,  {Schmidtobreick} L.,   {Trushkin} S.,  2011, \mn@doi
  [\mnras] {10.1111/j.1365-2966.2011.18785.x}, \href
  {http://adsabs.harvard.edu/abs/2011MNRAS.417.2401B} {417, 2401}

\bibitem[\protect\citeauthoryear{{Bowler}}{{Bowler}}{2010}]{2010A&A...521A..81B}
{Bowler} M.~G.,  2010, \mn@doi [\aap] {10.1051/0004-6361/201014711}, \href
  {http://adsabs.harvard.edu/abs/2010A%26A...521A..81B} {521, A81}

\bibitem[\protect\citeauthoryear{{Bowler}}{{Bowler}}{2011}]{2011A&A...531A.107B}
{Bowler} M.~G.,  2011, \mn@doi [\aap] {10.1051/0004-6361/201016381}, \href
  {http://adsabs.harvard.edu/abs/2011A%26A...531A.107B} {531, A107}

\bibitem[\protect\citeauthoryear{{Bowler}}{{Bowler}}{2018}]{2018A&A...619L...4B}
{Bowler} M.~G.,  2018, \mn@doi [\aap] {10.1051/0004-6361/201834121}, \href
  {http://adsabs.harvard.edu/abs/2018A%26A...619L...4B} {619, L4}

\bibitem[\protect\citeauthoryear{{Cherepashchuk}}{{Cherepashchuk}}{1981}]{1981MNRAS.194..761C}
{Cherepashchuk} A.~M.,  1981, \mn@doi [\mnras] {10.1093/mnras/194.3.761}, \href
  {http://adsabs.harvard.edu/abs/1981MNRAS.194..761C} {194, 761}

\bibitem[\protect\citeauthoryear{{Cherepashchuk}, {Aslanov}  \&
  {Kornilov}}{{Cherepashchuk} et~al.}{1982}]{1982SvA....26..697C}
{Cherepashchuk} A.~M.,  {Aslanov} A.~A.,   {Kornilov} V.~G.,  1982, \sovast,
  \href {http://adsabs.harvard.edu/abs/1982SvA....26..697C} {26, 697}

\bibitem[\protect\citeauthoryear{{Cherepashchuk}, {Esipov}, {Dodin}, {Davydov}
  \& {Belinskii}}{{Cherepashchuk} et~al.}{2018a}]{2018ARep...62..747C}
{Cherepashchuk} A.~M.,  {Esipov} V.~F.,  {Dodin} A.~V.,  {Davydov} V.~V.,
  {Belinskii} A.~A.,  2018a, \mn@doi [Astronomy Reports]
  {10.1134/S106377291811001X}, \href
  {http://adsabs.harvard.edu/abs/2018ARep...62..747C} {62, 747}

\bibitem[\protect\citeauthoryear{{Cherepashchuk}, {Postnov}, {Belinski}  \&
  {(Paper I)}}{{Cherepashchuk} et~al.}{2018b}]{2018MNRAS.479.4844C}
{Cherepashchuk} A.~M.,  {Postnov} K.~A.,  {Belinski} A.~A.,   {(Paper I)}
  2018b, \mn@doi [\mnras] {10.1093/mnras/sty1853}, \href
  {http://adsabs.harvard.edu/abs/2018MNRAS.479.4844C} {479, 4844}

\bibitem[\protect\citeauthoryear{{Fabrika}}{{Fabrika}}{2004}]{2004ASPRv..12....1F}
{Fabrika} S.,  2004, Astrophysics and Space Physics Reviews, \href
  {http://adsabs.harvard.edu/abs/2004ASPRv..12....1F} {12, 1}

\bibitem[\protect\citeauthoryear{{Goranskij}}{{Goranskij}}{2011}]{2011PZ.....31....5G}
{Goranskij} V.,  2011, Peremennye Zvezdy, \href
  {http://adsabs.harvard.edu/abs/2011PZ.....31....5G} {31}

\bibitem[\protect\citeauthoryear{{Gravity Collaboration} et~al.,}{{Gravity
  Collaboration} et~al.}{2017}]{2017A&A...602L..11G}
{Gravity Collaboration} et~al., 2017, \mn@doi [\aap]
  {10.1051/0004-6361/201731038}, \href
  {http://adsabs.harvard.edu/abs/2017A%26A...602L..11G} {602, L11}

\bibitem[\protect\citeauthoryear{{Hillwig} \& {Gies}}{{Hillwig} \&
  {Gies}}{2008}]{2008ApJ...676L..37H}
{Hillwig} T.~C.,  {Gies} D.~R.,  2008, \mn@doi [\apjl] {10.1086/587140}, \href
  {http://adsabs.harvard.edu/abs/2008ApJ...676L..37H} {676, L37}

\bibitem[\protect\citeauthoryear{{Hillwig}, {Gies}, {Huang}, {McSwain},
  {Stark}, {van der Meer}  \& {Kaper}}{{Hillwig}
  et~al.}{2004}]{2004ApJ...615..422H}
{Hillwig} T.~C.,  {Gies} D.~R.,  {Huang} W.,  {McSwain} M.~V.,  {Stark} M.~A.,
  {van der Meer} A.,   {Kaper} L.,  2004, \mn@doi [\apj] {10.1086/423927},
  \href {http://adsabs.harvard.edu/abs/2004ApJ...615..422H} {615, 422}

\bibitem[\protect\citeauthoryear{{Kawai}, {Matsuoka}, {Pan}  \&
  {Stewart}}{{Kawai} et~al.}{1989}]{1989PASJ...41..491K}
{Kawai} N.,  {Matsuoka} M.,  {Pan} H.-C.,   {Stewart} G.~C.,  1989, \pasj,
  \href {http://adsabs.harvard.edu/abs/1989PASJ...41..491K} {41, 491}

\bibitem[\protect\citeauthoryear{{Kotani}, {Kawai}, {Matsuoka}  \&
  {Brinkmann}}{{Kotani} et~al.}{1996}]{1996PASJ...48..619K}
{Kotani} T.,  {Kawai} N.,  {Matsuoka} M.,   {Brinkmann} W.,  1996, \mn@doi
  [\pasj] {10.1093/pasj/48.4.619}, \href
  {http://adsabs.harvard.edu/abs/1996PASJ...48..619K} {48, 619}

\bibitem[\protect\citeauthoryear{{Kubota}, {Ueda}, {Fabrika}, {Medvedev},
  {Barsukova}, {Sholukhova}  \& {Goranskij}}{{Kubota}
  et~al.}{2010}]{2010ApJ...709.1374K}
{Kubota} K.,  {Ueda} Y.,  {Fabrika} S.,  {Medvedev} A.,  {Barsukova} E.~A.,
  {Sholukhova} O.,   {Goranskij} V.~P.,  2010, \mn@doi [\apj]
  {10.1088/0004-637X/709/2/1374}, \href
  {http://adsabs.harvard.edu/abs/2010ApJ...709.1374K} {709, 1374}

\bibitem[\protect\citeauthoryear{{Margon}}{{Margon}}{1984}]{1984ARA&A..22..507M}
{Margon} B.,  1984, \mn@doi [\araa] {10.1146/annurev.aa.22.090184.002451},
  \href {http://adsabs.harvard.edu/abs/1984ARA%26A..22..507M} {22, 507}

\bibitem[\protect\citeauthoryear{{Margon}, {Ford}, {Grandi}  \&
  {Stone}}{{Margon} et~al.}{1979}]{1979ApJ...233L..63M}
{Margon} B.,  {Ford} H.~C.,  {Grandi} S.~A.,   {Stone} R.~P.~S.,  1979, \mn@doi
  [\apjl] {10.1086/183077}, \href
  {http://adsabs.harvard.edu/abs/1979ApJ...233L..63M} {233, L63}

\bibitem[\protect\citeauthoryear{{Massevich} \& {Tututkov}}{{Massevich} \&
  {Tututkov}}{1988}]{MT1988}
{Massevich} A.~G.,  {Tututkov} A.~V.,  1988, Evolution of Stars: Theory and
  Observations.
Moscow: Nauka

\bibitem[\protect\citeauthoryear{{Pavlovskii} \& {Ivanova}}{{Pavlovskii} \&
  {Ivanova}}{2015}]{2015MNRAS.449.4415P}
{Pavlovskii} K.,  {Ivanova} N.,  2015, \mn@doi [\mnras] {10.1093/mnras/stv619},
  \href {http://adsabs.harvard.edu/abs/2015MNRAS.449.4415P} {449, 4415}

\bibitem[\protect\citeauthoryear{{Pavlovskii}, {Ivanova}, {Belczynski}  \&
  {Van}}{{Pavlovskii} et~al.}{2017}]{2017MNRAS.465.2092P}
{Pavlovskii} K.,  {Ivanova} N.,  {Belczynski} K.,   {Van} K.~X.,  2017, \mn@doi
  [\mnras] {10.1093/mnras/stw2786}, \href
  {http://adsabs.harvard.edu/abs/2017MNRAS.465.2092P} {465, 2092}

\bibitem[\protect\citeauthoryear{{Robinson}, {Froning}, {Jaffe}, {Kaplan},
  {Kim}, {Mace}, {Sokal}  \& {Lee}}{{Robinson}
  et~al.}{2017}]{2017ApJ...841...79R}
{Robinson} E.~L.,  {Froning} C.~S.,  {Jaffe} D.~T.,  {Kaplan} K.~F.,  {Kim} H.,
   {Mace} G.~N.,  {Sokal} K.~R.,   {Lee} J.-J.,  2017, \mn@doi [\apj]
  {10.3847/1538-4357/aa6f0c}, \href
  {http://adsabs.harvard.edu/abs/2017ApJ...841...79R} {841, 79}

\bibitem[\protect\citeauthoryear{{Soberman}, {Phinney}  \& {van den
  Heuvel}}{{Soberman} et~al.}{1997}]{1997A&A...327..620S}
{Soberman} G.~E.,  {Phinney} E.~S.,   {van den Heuvel} E.~P.~J.,  1997, \aap,
  \href {http://adsabs.harvard.edu/abs/1997A%26A...327..620S} {327, 620}

\bibitem[\protect\citeauthoryear{{Waisberg}, {Dexter}, {Olivier-Petrucci},
  {Dubus}  \& {Perraut}}{{Waisberg} et~al.}{2018}]{2018arXiv181112558W}
{Waisberg} I.,  {Dexter} J.,  {Olivier-Petrucci} P.,  {Dubus} G.,   {Perraut}
  K.,  2018, arXiv e-prints, \href
  {http://adsabs.harvard.edu/abs/2018arXiv181112558W} {}

\bibitem[\protect\citeauthoryear{{van den Heuvel}, {Portegies Zwart}  \& {de
  Mink}}{{van den Heuvel} et~al.}{2017}]{2017MNRAS.471.4256V}
{van den Heuvel} E.~P.~J.,  {Portegies Zwart} S.~F.,   {de Mink} S.~E.,  2017,
  \mn@doi [\mnras] {10.1093/mnras/stx1430}, \href
  {http://adsabs.harvard.edu/abs/2017MNRAS.471.4256V} {471, 4256}

\makeatother
\end{thebibliography}

\bsp	
\label{lastpage}
\end{document}